\documentclass[10pt,twocolumn,letterpaper]{article}
\pdfoutput=1 
\usepackage{iccv}              
\usepackage{multirow}
\usepackage{caption}
\usepackage{subcaption}
\expandafter\def\csname ver@subfig.sty\endcsname{}
\usepackage{graphicx,subfig}
\usepackage[accsupp]{axessibility}  
\usepackage{latexsym}

\usepackage{color}
\definecolor{green}{rgb}{0, 0.5, 0}
\definecolor{orange}{rgb}{0.8, 0.6, 0.2}
\definecolor{orange2}{rgb}{1.0, 0.6, 0.2}
\definecolor{red}{rgb}{1.0, 0.0, 0.0}
\definecolor{teal}{rgb}{0.0, 0.4, 0.4}
\definecolor{purple}{rgb}{0.65,0,0.65}
\definecolor{saffron}{rgb}{0.95,0.75,0.2}
\definecolor{turquoise}{rgb}{0.0,0.5,0.5}
\definecolor{black}{rgb}{0.0, 0.0, 0.0}
\definecolor{gray}{rgb}{0.5, 0.5, 0.5}
\definecolor{softblue}{RGB}{0,122,204}

\newcommand{\mzh}[1]{{\color{black}#1}}

\definecolor{iccvblue}{rgb}{0.21,0.49,0.74}
\usepackage[pagebackref,breaklinks,colorlinks,allcolors=iccvblue]{hyperref}

\title{TAG-WM: Tamper-Aware Generative Image Watermarking via Diffusion Inversion Sensitivity}

\author{
Yuzhuo Chen$^{1}$ \quad Zehua Ma$^{1}\thanks{Corresponding author}$ \quad Han Fang$^{2}$ \quad Weiming Zhang$^{1}$ \quad Nenghai Yu$^{1}$\\
$^{1}$Anhui Province Key Laboratory of Digital Security, University of Science and Technology of China \\
$^{2}$School of Computing, National University of Singapore \\
{\tt\small $^{1}$yz.chen@mail.ustc.edu.cn;
$^{1}$\{mzh045, zhangwm, ynh\}@ustc.edu.cn;
$^{2}$fanghan@nus.edu.sg
}}

\begin{document}
\maketitle

\begin{abstract} 
\label{Abstract}
AI-generated content (AIGC) enables efficient visual creation but raises copyright and authenticity risks. As a common technique for integrity verification and source tracing, digital image watermarking is regarded as a potential solution to above issues. However, the widespread adoption and advancing capabilities of generative image editing tools have amplified malicious tampering risks, while simultaneously posing new challenges to passive tampering detection and watermark robustness. To address these challenges, this paper proposes a \textbf{T}amper-\textbf{A}ware \textbf{G}enerative image \textbf{W}ater\textbf{M}arking method named TAG-WM. The proposed method comprises four key modules: a dual-mark joint sampling (DMJS) algorithm for embedding copyright and localization watermarks into the latent space while preserving generative quality, the watermark latent reconstruction (WLR) utilizing reversed DMJS, a dense variation region detector (DVRD) leveraging diffusion inversion sensitivity to identify tampered areas via statistical deviation analysis, and the tamper-aware decoding (TAD) guided by localization results. The experimental results demonstrate that TAG-WM achieves state-of-the-art performance in both tampering robustness and localization capability even under distortion, while preserving lossless generation quality and maintaining a watermark capacity of 256 bits. The code is available at: {\normalfont\href{https://github.com/Suchenl/TAG-WM}{\textcolor{softblue}{\textit{\underline{https://github.com/Suchenl/TAG-WM}}}}}.
\end{abstract}

\section{Introduction}
\label{Introduction}

The rise of AI-generated content (AIGC) has garnered significant attention across various fields, creating substantial commercial value. Particularly in visual content generation, the advent of diffusion models \cite{DMs-Imagen3,DMs-DALLE3,DMs-LDMs} has sparked the emergence of numerous image generation and manipulation applications \cite{DMs-ControlNet}, enabling individuals across industries to easily and efficiently perform customized generation or editing of images with high quality. However, this technological accessibility inevitably leads to uncertainties in the provenance and authenticity of images in the AIGC era, resulting in copyright risks and disinformation threats. For instance, users might use image generators to create similar graphics to copyrighted work and falsely claim ownership. Furthermore, with image generators, malicious actors could easily create or manipulate lifelike images to spread false information about non-existent events.

As a common technique for integrity verification and source tracing, digital image watermarking has gained increasing prominence in AIGC. \mzh{Existing image watermarking methods were primarily designed by traditional image processing algorithms \cite{old_wm-wavelet,old_wm-combined_DWT_DCT} or deep neural networks \cite{postprocessing_watermarking-MBRS,postprocessing_watermarking-PIMoG}, both of which embed robust watermarks through post-processing cover images. However, such post-processing embedding approaches inevitably introduce visual artifacts. Considering that the quality of generated images is the goal pursued by image generation models at a high computational cost, such watermarking methods contradict the application scenarios of generative models. With the rise of diffusion models, new watermarking paradigms have emerged to protect AI-generated content. Recent advancements focus on model fine-tuning \cite{fine_tuning_based_watermarking-Diffusionshield,fine_tuning_based_watermarking-StableSignature,fine_tuning_based_watermarking-flexible} and latent space-based watermarks \cite{latent_representation_based_watermarking-Treering}, watermarking directly during the generative process itself. Such watermarking methods embedded in the generative process can better integrate watermark information with image content while significantly reducing the impact of watermark embedding on the visual quality of generated images. The Gaussian Shading (GS) proposed by Yang et al. \cite{gaussianshading} takes a significant step forward, achieving provably visual quality lossless watermarking for generated images. By employing distribution-preserving sampling to map watermarks into latent space representations, GS ensures the watermarked latent becomes statistically indistinguishable from the original one, thereby maintaining the visual quality equivalent to the original generative images.
}

\mzh{Meanwhile, the emergence of generative models like ControlNet \cite{DMs-ControlNet}, which enable high-quality image modifications, has exacerbated the threat of malicious tampering, imposing stricter requirements for watermark robustness and tamper localization capabilities. Unlike common image distortions, malicious alterations fundamentally alter pixel values. Unfortunately, the DDIM inversion process demonstrates sensitivity to pixel modifications (Sec. \ref{sec:smp_ddim}), rendering inversion-based generative watermarking methods like GS less robust against image manipulation distortions (e.g., cropping, tampering). On the other hand, as generative model-based image manipulations become increasingly indistinguishable, applying watermarks to achieve more generalizable proactive tamper localization has emerged as a novel and practical demand in watermark functionality. For instance, MaLP \cite{tl-proactive-MaLP} achieves tamper detection and pixel-level localization through learned template embedding. EditGuard \cite{wm_and_tl-Editguard} simultaneously embeds both identification watermarks and tampering localization watermarks into images, decoding them in parallel to accomplish dual objectives of copyright verification and tamper localization.}

\mzh{To address the aforementioned challenges while introducing tamper localization capabilities, we propose a tamper-aware generative image watermarking named TAG-WM. This framework leverages the sensitivity of DDIM inversion to pixel modifications to design tamper localization strategies, and further employs the localization results as confidence guidance for watermark decoding, thereby enhancing the accuracy of watermark extraction under tampering distortions.}


The main contributions of this paper are as follows:
\begin{itemize}
    \item We propose a dual-mark joint sampling algorithm to simultaneously embed copyright watermark and localization watermark into the latent space of diffusion models. This strategy preserves standard normal distributions of latent representation, ensuring lossless visual quality while introducing tampering localization capability.
    
    \item We develop a dense variation region detector for tampering localization, leveraging the sensitivity of diffusion inversion to image modifications. By analyzing statistical deviations between original/reconstructed localization watermarks, the proposed detection method demonstrates strong generalization capability.

    \item We introduce tamper-aware message decoding guided by tampering localization results, which improves the robustness of such generated image watermarking methods against image modifications.
\end{itemize}


\section{Related Work}
\label{Related Work}

\noindent\textbf{Diffusion-based Image Generation}
The explosive growth of diffusion models has revolutionized image synthesis, with frameworks like DDPM \cite{DMs-DDPMs} enabling high-fidelity generation, DDIM \cite{DMs-scheduler-DDIM} enabling fast deterministic sampling via non-Markov processes, LDMs \cite{DMs-LDMs} compressing data into latent space for efficiency, text-guided models like Stable Diffusion and Imagen \cite{DMs-Imagen3} leveraging large-scale pretraining for diverse generations, while ControlNet \cite{DMs-ControlNet} facilitates controllable creation and manipulation. These advancements have made diffusion models the standard for generative tasks, yet their very success intensifies the urgency of addressing copyright risks in generated content.

\noindent\textbf{Watermarking for Generative Models}
While post-processing methods \cite{postprocessing_watermarking-HiDDeN,postprocessing_watermarking-REVMark, postprocessing_watermarking-SepMark} suffer from capacity-quality trade-offs, Tree-Ring \cite{latent_representation_based_watermarking-Treering} proposed \textit{in-generation} approaches for popular DMs, which bases on the non-Markov process and deterministic of DDIM inversion \cite{DMs-scheduler-DDIM}. Gaussian Shading \cite{gaussianshading} uses it and achieves provable quality-lossless multi-bit watermarking. However, their reliance on noise patterns makes them vulnerable to image manipulations — a critical limitation given the prevalence of image editing tools. This motivates our tamper-aware watermarking that inherits in-generation advantages while resisting post-hoc manipulations. 

\noindent\textbf{Tampering Localization}
Passive methods detect specific manipulation traces \cite{tl-passive-specific-OSN} or general artifacts \cite{tl-passive-universal-MVSSNet}. As diffusion-based manipulation \cite{ImageInpainting-SDXL, DMs-ControlNet} increasingly produces undetectable forgeries, passive detectors face diminishing returns. Proactive solutions like EditGuard \cite{wm_and_tl-Editguard} circumvent this issue by employing watermarking techniques, but they suffer from generation-time overheads, image quality degradation, and limited robustness against out-of-distribution (OOD) samples (e.g., untrained degraded inputs). Our approach addresses these limitations by embedding localization watermarks directly into the diffusion pipeline, ensuring seamless integration without compromising efficiency or fidelity.


\begin{figure*}[t]
  \centering
  \includegraphics[width=1\linewidth]{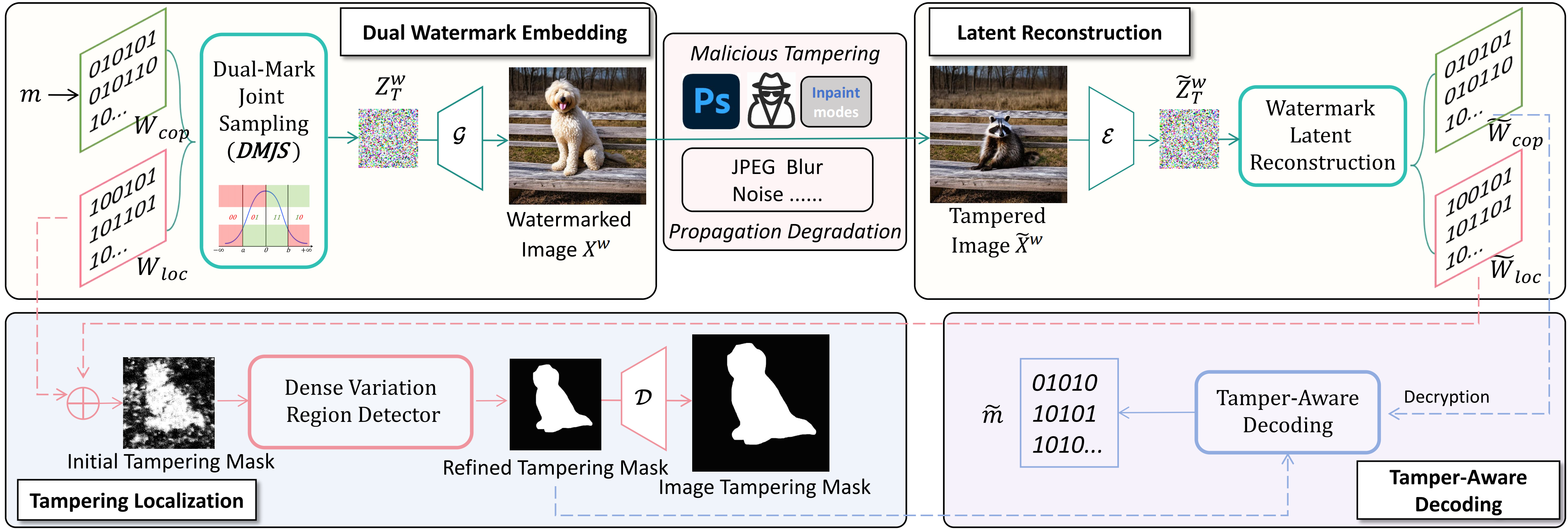} 
    \caption{The proposed TAG-WM framework. It embeds copyright $W_{cop}$ and localization watermarks $W_{loc}$ through dual-watermark joint sampling strategy. By analyzing dense variation regions of $W_{loc}$, it enables tamper localization while improving watermark decoding accuracy using tampering insights.
}
\label{fig: Framework of our method} 
\end{figure*}

\section{Preliminaries}
\label{sec:Preliminaries}
\subsection{Denoising Diffusion Implicit Model}
The Denoising Diffusion Implicit Model (DDIM) accelerates diffusion-based generation by defining a non-Markovian process. It enables deterministic sampling through a modified reverse process. The generation process can be formulated as:
\begin{equation}
\scalebox{0.9}{$
    x_{t-1} = \sqrt{\frac{\bar\alpha_{t-1}}{\bar\alpha_t}}x_t+(\sqrt{1-\bar\alpha_{t-1}}-\sqrt{\frac{\bar\alpha_{t-1}(1-\bar\alpha_{t})}{\bar\alpha_{t}}})\epsilon_{\theta}^{(t)}(x_t)
    $}
\end{equation}
Suppose $\epsilon_{\theta}^{(t)}(x_t)\approx\epsilon_{\theta}^{(t-1)}(x_{t-1})$, then the inversion process is as follows:
\begin{equation}
\label{eq:DDIM_inversion}
\scalebox{0.87}{$
    x_{t} \approx \sqrt{\frac{\bar\alpha_{t}}{\bar\alpha_{t-1}}}x_{t-1}+(\sqrt{1-\bar\alpha_{t}}-\sqrt{\frac{\bar\alpha_{t}(1-\bar\alpha_{t-1})}{\bar\alpha_{t-1}}})\epsilon_{\theta}^{(t-1)}(x_{t-1})
    $}
\end{equation}
which we used to recover watermarked pseudo-noise.

\subsection{Spatial Mapping Properties}
\label{sec:smp_ddim}
According to the mapping relationship Eq.~(\ref{eq:DDIM_inversion}) between the noise $x_t$ and the image $x_0$ in the DDIM framework, it is evident that any modification to $x_0$ in pixel space will propagate to the corresponding positions in $x_t$. This property is preserved to some extent in the VAE-based latent-diffusion architecture \cite{DMs-LDMs}, as the VAE retains spatial structure while accelerating inference. Consequently, spatial perturbations in the image domain induce localized changes in the latent space during inversion. 

\section{Method}
\label{Method}


As illustrated in Fig. \ref{fig: Framework of our method}, the proposed TAG-WM framework operates through four stages. During watermark embedding, our method encodes both copyright watermark $W_{cop}$ and localization watermark $W_{loc}$ via a dual-watermark joint sampling strategy integrated into the diffusion model generation process. For watermark extraction, the framework first reconstructs latent watermark components through inverse sampling. Subsequently, it performs tampering localization by analyzing dense variation regions in $W_{loc}$, while simultaneously leveraging this tampering information to enhance the decoding accuracy of copyright messages.

\subsection{Dual Watermark Embedding}
\paragraph{Watermark Initialization}
The watermark generation process involves creating both copyright watermark $W_{cop}$ and localization watermark $W_{loc}$. For $W_{cop}$, the copyright messages $m \in \{0,1\}^L$ are adaptively expanded to match the dimensional space of the latent representations in generation models. For latent variables with dimensions $C \times H \times W$, the expansion process operates as follows:
\begin{equation}
\hat{m} = \Bigl[\, \overbrace{m, \dots, m}^{N},\, m_{1:R} \,\Bigr]
\end{equation}
where $N = \lfloor D/L \rfloor$ represents full replication count and $R = D \bmod L$ corresponds to residual bits, $D =C \times H \times W$. Then, the expanded message $\hat{m}$ is encrypted via ChaCha20 \cite{chacha20} with key $k$ to obtain $W_{cop}$ with a multiple binary uniform distribution:
\begin{equation}
W_{cop} = \mathrm{Reshape}\left(\mathrm{ChaCha20}(\hat{m}, k), (C,H,W)\right)
\end{equation}

For the localization watermark $W_{loc}$, a deterministic pseudo-random process is employed. Using a fixed seed $s$, a pseudo-random generator $G$ is adopted to generate $W_{loc}$:
\begin{equation}
W_{loc} = \mathrm{Reshape}\left(G(s, \theta),\ (C,H,W)\right)
\end{equation}
where each element follows $\mathcal{B}(1-\theta)$. $\mathcal{B}$ donates the Bernoulli distribution, and the $\theta$ is the probability of zeros.

\begin{figure}[t]
    \centering
    \begin{subfigure}[b]{0.23\textwidth}
        \includegraphics[width=\linewidth]{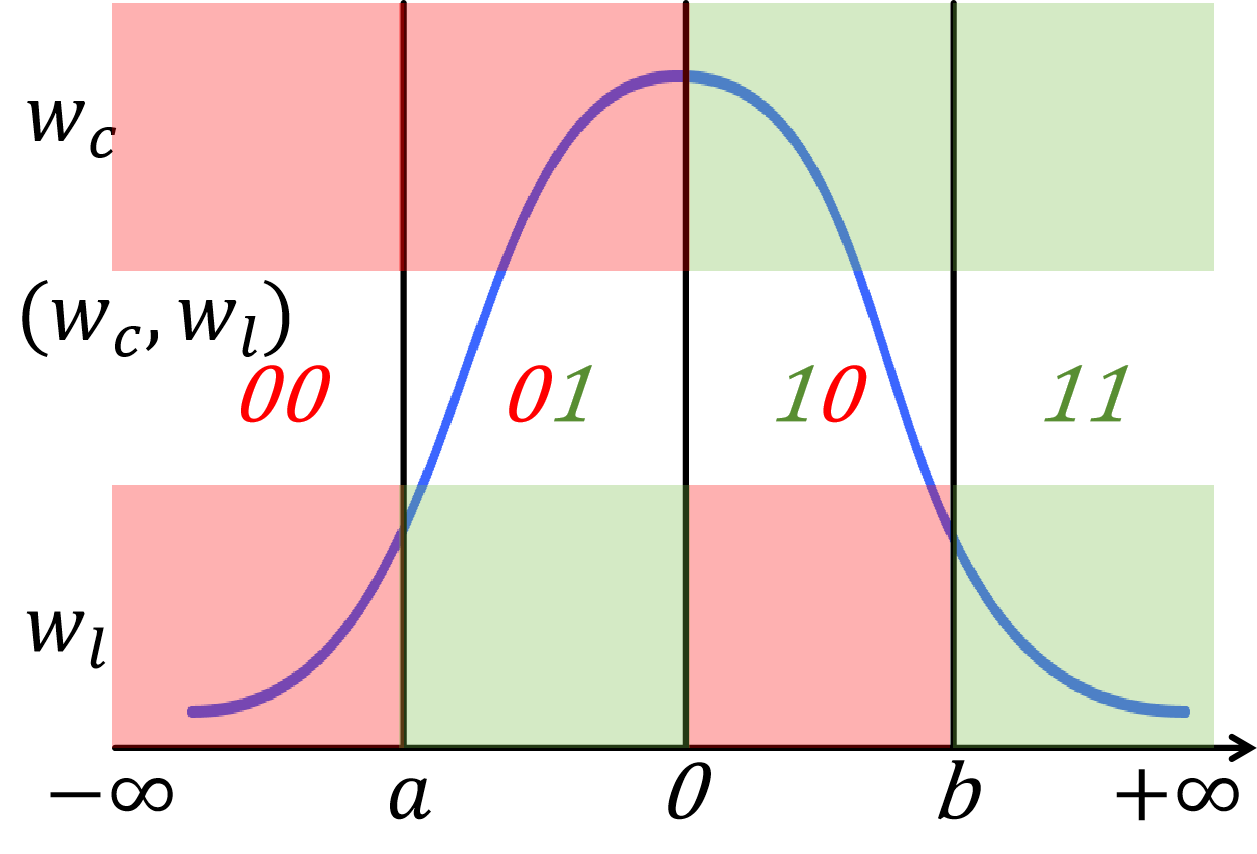}
        \caption{Four-interval strategy}
    \end{subfigure}
    \begin{subfigure}[b]{0.23\textwidth}
        \includegraphics[width=\linewidth]{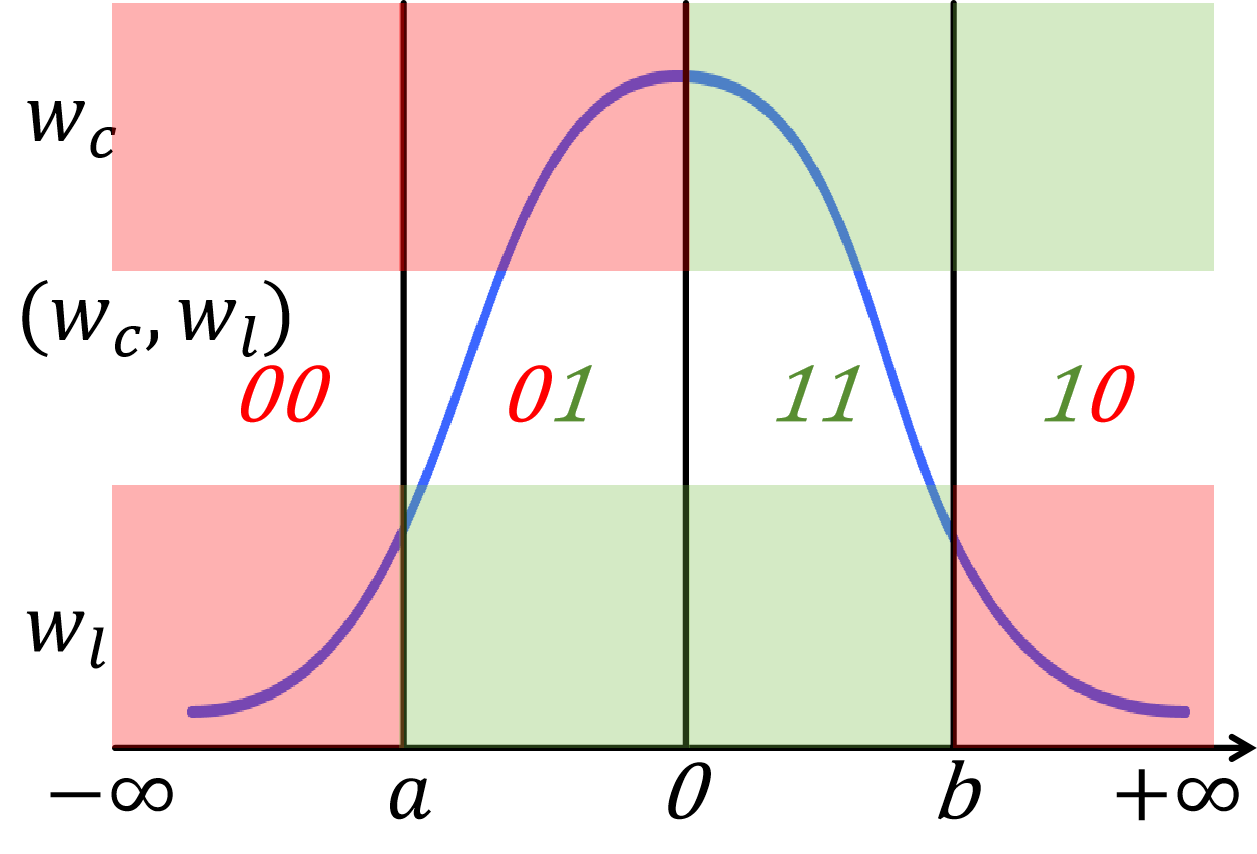}
        \caption{Three-interval strategy}
    \end{subfigure}
    \caption{Embedding strategies for bit pairs $(w_{c},w_{l})$.}
    \label{fig:embedding_orders}
\end{figure}

\paragraph{Dual-Mark Joint Sampling}
\label{sec:dmjs} 
To embed dual watermarks while preserving visual quality, we propose the Dual-Mark Joint Sampling (DMJS) algorithm, which generates dual-watermarked pseudo-Gaussian noise $Z_T^w \sim \mathcal{N}(\boldsymbol{0},\boldsymbol{I})$ according to $W_{cop}$ and $W_{loc}$. Each element $z$ in $Z_T^w$ corresponds to the bit combination $(w_c, w_l) \in \{(0,0),(0,1),(1,0),(1,1)\}$ of $W_{cop}$ and $W_{loc}$, with its distribution expressed as:
\begin{equation}
p(z) = \sum_{(w_c,w_l) \in \{0,1\}^2} P(w_c,w_l) \cdot p(z|w_c,w_l)
\end{equation}
Then, the real number set $\mathbb{R}$ can be partitioned into four continuous subsets for watermark embedding: $(-\infty, a)$, $[a,0)$, $[0,b)$, $[b,+\infty)$. A straightforward embedding idea is to create a direct mapping of these four intervals to the bit combination $(0,0),(0,1),(1,0),(1,1)$. However, to mitigate boundary errors from DDIM inversion and noise prediction, we reduce the intervals to three: $(-\infty, a)$, $[a,b)$, $[b,+\infty)$ by reordering the embedding sequence to $(0,0),(1,0),(1,1),(1,0)$. This retains two intervals for $W_{cop}$ and three for $W_{loc}$; see Fig. \ref{fig:embedding_orders}.

Both types of strategies can ensure $Z_T^w \sim \mathcal{N}(\boldsymbol{0},\boldsymbol{I})$, thereby preserving the generation quality without degradation. The detailed proof is as follows. The conditional density for $Z_T^w$ is:
\begin{equation}
p(z|w_{c},w_{l}) = \frac{\phi(z)}{\mid\Phi(l_{w_{c\&l}})-\Phi(u_{w_{c\&l}})\mid} \cdot \mathbf{1}_{[l_{w_{c\&l}}, u_{w_{c\&l}})}(z)
\end{equation}
where $\phi(\cdot)$ and $\Phi(\cdot)$ are the PDF and CDF of $\mathcal{N}(0,1)$, respectively. $\mathbf{1}_{[a, b)}(\cdot)$ denotes the indicator function, which equals 1 if
$z$ lies in the interval $[a, b)$ and 0 otherwise. The interval lower boundary  $l_{w_{c\&l}}$ and the interval upper boundary $u_{w_{c\&l}}$ are strategy-dependent as shown in 
\ref{table: Sampling boundaries}, in which $a$ and $b$ are defined as:
\begin{equation}
a = \Phi^{-1}(\tfrac{\theta}{2}), b = \begin{cases} 
\Phi^{-1}(\tfrac{1}{2} + \tfrac{\theta}{2}) & \text{(four intervals)} \\
\Phi^{-1}(1 - \tfrac{\theta}{2}) & \text{(three intervals)}
\end{cases}
\end{equation}
So the probability distribution of $z$ can be denoted as:
\begin{equation}
\resizebox{0.423\textwidth}{!}{$
\begin{aligned}
p(z) 
&= \sum_{(w_c,w_l) \in \{0,1\}^2} \theta^{1-w_{l}}(1-\theta)^{w_{l}} \cdot \tfrac{1}{2} \cdot p(z|w_{c},w_{l})\\
&= \phi(z)
\end{aligned}
$}
\end{equation}
confirming $Z_T^w$ maintains the standard normal distribution required by diffusion models.

Finally, we employ Ordinary Differential Equations (ODEs) \cite{DMs-scheduler-DDIM} to denoise $Z_T^w$ to obtain the denoised watermarked latent $Z_0^w$, which is subsequently decoded by the VAE to generate the final watermarked image $X^w$.

\subsection{Watermark Latent Reconstruction}
In the extraction, we first reconstruct watermark latents from a tampered and degraded image ($\widetilde{X}^{w}$). To restore the sampled noise, we first use VAE to encode $\widetilde{X}^{w}$ into latent space, resulting in $\widetilde{Z}_0^{w}$. Next, we apply DDIM inversion to obtain the restored watermarked pseudo-Gaussian noise $\widetilde{Z}_T^w$, which resembles the sampled noise $Z_T^w$. We then reverse the sampling process according to the boundaries specified in Table \ref{table: Sampling boundaries}. Based on the number of intervals chosen and the boundaries associated with each reversed bit, we can reconstruct the watermarks, denoted as $\widetilde{W}_{cop}$ and $\widetilde{W}_{loc}$.




\begin{table}[t]
\centering
\resizebox{\linewidth}{!}{
\setlength{\tabcolsep}{.3in}{
\begin{tabular}{ccc}
\toprule
\multirow{2}{*}{$(w_c, w_l)$} & \multicolumn{2}{c}{Boundaries $(l_{w_{c\&l}}, u_{w_{c\&l}})$} \\ \cmidrule(r){2-3}
& Four intervals & Three intervals \\
\midrule
$(0,0)$ & $(-\infty,a)$ & $(-\infty,a)$ \\
$(0,1)$ & $[a, 0)$ & $[a, 0)$ \\
$(1,0)$ & $[0,b)$ & $[b, +\infty)$ \\
$(1,1)$ & $[b, +\infty)$ & $[0,b)$ \\
\bottomrule
\end{tabular}
}}
\caption{Sampling boundaries under different strategies.}
\label{table: Sampling boundaries}
\end{table}

\subsection{Tampering Localization}
\label{Tampering Localization}
The proposed TAG-WM localizes the tampered regions using the reconstructed localization watermark $\widetilde{W}_{loc}$ and the original $W_{loc}$ generated by the shared seed $s$. First, the initial tampering estimation $\widetilde{M}_{tam}^{ini}$ can be calculated through XOR operation between $W_{loc}$ and $\widetilde{W}_{loc}$. To refine the localization result, we propose a \textbf{d}ense \textbf{v}ariation \textbf{r}egion \textbf{d}etector (DVRD), motivated by the observation that local tampering disrupts the deterministic relationship between $W_{loc}$ and $\widetilde{W}_{loc}$, thereby randomizing the bit distribution at latent space positions corresponding to the tampered regions.

This randomization process can be modeled as follows. Let event $A$ denote tampering in the image region corresponding to $\widetilde{w}_{l}\  (\in \widetilde{W}_{loc})$, defined as:
\[
A = \left\{ \widetilde{w}_{l} \, \middle| \, \parbox{0.35\textwidth}{The image region corresponding to \( \widetilde{w}_{l} \) is tampered.} \right\}
\]
Then, the conditional probabilities are expressed as:
\begin{equation}
P(\widetilde{w}_{l}=0\mid A)=\theta, \quad P(\widetilde{w}_{l}=1\mid A)=1-\theta
\end{equation}
The error probability under tampering becomes:
\begin{equation}
P(\widetilde{w}_{l}\neq w_{l}\mid A) = 2\theta(1-\theta)
\end{equation}
which is a concave function that peaks at 0.5 when $\theta=0.5$. Setting $\theta=0.5$, the intrinsic error probability $P(\widetilde{w}_{l}\neq w_{l}\mid \bar{A})$ is experimentally measured as 0.14513. The significant gap between maximum tampering-induced errors (0.5) and intrinsic errors (0.14513) enables threshold-based detection through analysis of dense variation region in $\widetilde{M}_{tam}^{ini}$.

The DVRD algorithm has two implementations:  
1) \underline{\textit{Train-free DVRD}} computes densities of variation region via multi-scale convolution kernels to balance fine/coarse detection. However, manual kernel optimization proves impractical due to infinite kernel size combinations. 
2) \underline{\textit{Trainable DVRD}} addresses this limitation using a UNet architecture that automatically learns optimal thresholds and weights across scales through downsampling/upsampling operations and skip connections. By training the detection model, the proposed method can refine coarse tamper localization masks $\widetilde{M}_{tam}^{ini}$ into latent tensor $\widetilde{M}_{tam}^{ref}$ that more accurately aligns with the image tampering regions without manual configuration.

Training methodology and comparative performance analysis between both DVRD variants are detailed in Sec~\ref{Training Details for Trainable DVRD} and Sec~\ref{Impact of Training and Number of Intervals}, respectively. The final output is the refined tampering localization mask $\widetilde{M}_{tam}^{ref}$. To get the binary tampering localization mask $\widetilde{M}_{tam}$ in image space, we decode $\widetilde{M}_{tam}^{ref}$ through the VAE decoder and binarize the decoded output by applying a threshold of 0.

\subsection{Tamper-Aware Copyright Message Decoding}
\label{Tamper-aware Copyright Message Extraction}
Finally, we decode copyright messages from $\widetilde{W}_{cop}$, using the refined tampering mask $\widetilde{M}_{tam}^{ref}$ as a guide to enhance robustness against tampering distortions. Specifically, we first decrypt $\widetilde{W}_{cop}$ with the original encryption key and then apply \textbf{t}amper-\textbf{a}ware \textbf{d}ecoding (TAD). By excluding watermark bits $w_c$ identified as compromised in $\widetilde{M}_{tam}^{ref}$, the system aggregates remaining reliable bits through majority voting to reconstruct $\widetilde{m}$. This selective exclusion of tampered regions significantly improves copyright message recovery accuracy under malicious modifications.




\section{Experiments}
\label{Experiments}
\subsection{Implementation Details}
\noindent\textbf{SD Models}.
In our experiments, we employed a text-to-image latent diffusion model (LDM) and chose Stable Diffusion (SD) from Hugging Face as our implementation. We evaluate TAG-WM as well as baseline methods, using three versions of SD: V1.4, V2.0, and V2.1. The size of the generated images is $512\times 512$, and the latent space dimension is $4\times 64\times 64$. 

\noindent\textbf{Benckmark Dataset}.
During inference, we employ the prompt from Stable-Diffusion-Prompt\footnote{https://huggingface.co/datasets/Gustavosta/Stable-Diffusion-Prompts}, with a guidance scale of 7.5. We sample 50 steps using DPMSolver \cite{DMs-scheduler-DPMSolver}. Considering that users tend to propagate the generated images without retaining the corresponding prompts, we use an empty prompt for inversion, with a scale of 1. We perform 50 steps of inversion using DDIM inversion \cite{DMs-scheduler-DDIM}. Through that, we get our test set includes 1,000 images, which are kept completely isolated from the training and validation sets. All degraded data samples were randomly selected from the predefined degradation types and intensity levels specified in Appendix \hyperref[sup: Implementation of Random Image Degradations]{A}.


\noindent\textbf{Final Settings}. 
In the main experiments, we set the $\theta=0.5$, the number of tampering localization template intervals$=3$, the DVRD to the trainable one, and the capacity of the copyright watermark to 256 bits. 



\subsection{Watermark Performance}

\begin{table*}[htbp]
\centering
\resizebox{\textwidth}{!}{
\begin{tabular}{cccccccccccc}
\toprule
\textbf{Methods} & \textbf{T-TPR (Clean)} & \textbf{T-TPR (Degraded)} & \textbf{Bit Acc (Clean)} & \textbf{Bit Acc (Degraded)} & \textbf{CLIP Score (t-value)} \\ 
\midrule
Stable Diffusion  &  -  &  -  &  -  &  -  & 0.3629$\pm$.0006 \\ \hline
DwtDct & 0.825/0.881/0.866 & 0.172/0.178/0.173 & 0.8030/0.8059/0.8023 & 0.5696/0.5671/0.5622 & 0.3617$\pm$.0007 (3.045) \\ 
DwtDctSvd & \textbf{1.000/1.000/1.000} & 0.597/0.594/0.599 & 0.9997/0.9987/0.9987 & 0.6920/0.6868/0.6905 & 0.3609$\pm$.0009 (4.452) \\ 
RivaGAN & 0.920/0.945/0.963 & 0.697/0.697/0.706 & 0.9762/0.9877/0.9921 & 0.8986/0.9124/0.9019 & 0.3611$\pm$.0009 (4.259) \\ 
EditGuard & \textbf{1.000/1.000/1.000} & 0.522/0.520/0.524 & 0.9999/0.9998/0.9998 & 0.7835/0.7839/0.7838 & 0.3621$\pm$.0027 (4.864)\\ 
Stable Signature & \textbf{1.000/1.000/1.000} & 0.502/0.505/0.496 & 0.9987/0.9978/0.9979 & 0.7520/0.7472/0.7500 & 0.3622$\pm$.0027 (0.7066)\\ 
Tree-Ring & \textbf{1.000/1.000/1.000} & 0.894/0.898/0.906 & - & - & 0.3632$\pm$.0006 (0.8278)\\ 
Gaussian Shading & \textbf{1.000/1.000/1.000} & 0.997/0.998/0.996 & \textbf{0.9999/0.9999/0.9999} & 0.9753/0.9749/0.9724 & 
0.3631$\pm$.0005 (\textbf{0.6870})\\ 
\textbf{TAG-WM (ours)} & \textbf{1.000/1.000/1.000} &  
\textbf{0.998/0.999/0.997}  &  \textbf{0.9999/0.9999/0.9999}  & \textbf{0.9756/0.9753/0.9726}  & 
0.3631$\pm$.0005 (\textbf{0.6870})\\ 
\bottomrule
\end{tabular}
}
\caption{Comparative results in non-tampering scenarios with baseline methods. We control the FPR at $10^{-6}$, and evaluate the T-TPR and bit accuracy for SD V1.4/V2.0/V2.1. To assess the bias in model performance, we conduct a t-test on SD V2.1.}
\label{table: Overall comparative results with watermarking methods}
\end{table*}

\begin{figure*}[htbp]
    \centering
    \begin{subfigure}[t]{0.246\textwidth}
        \includegraphics[width=\linewidth]{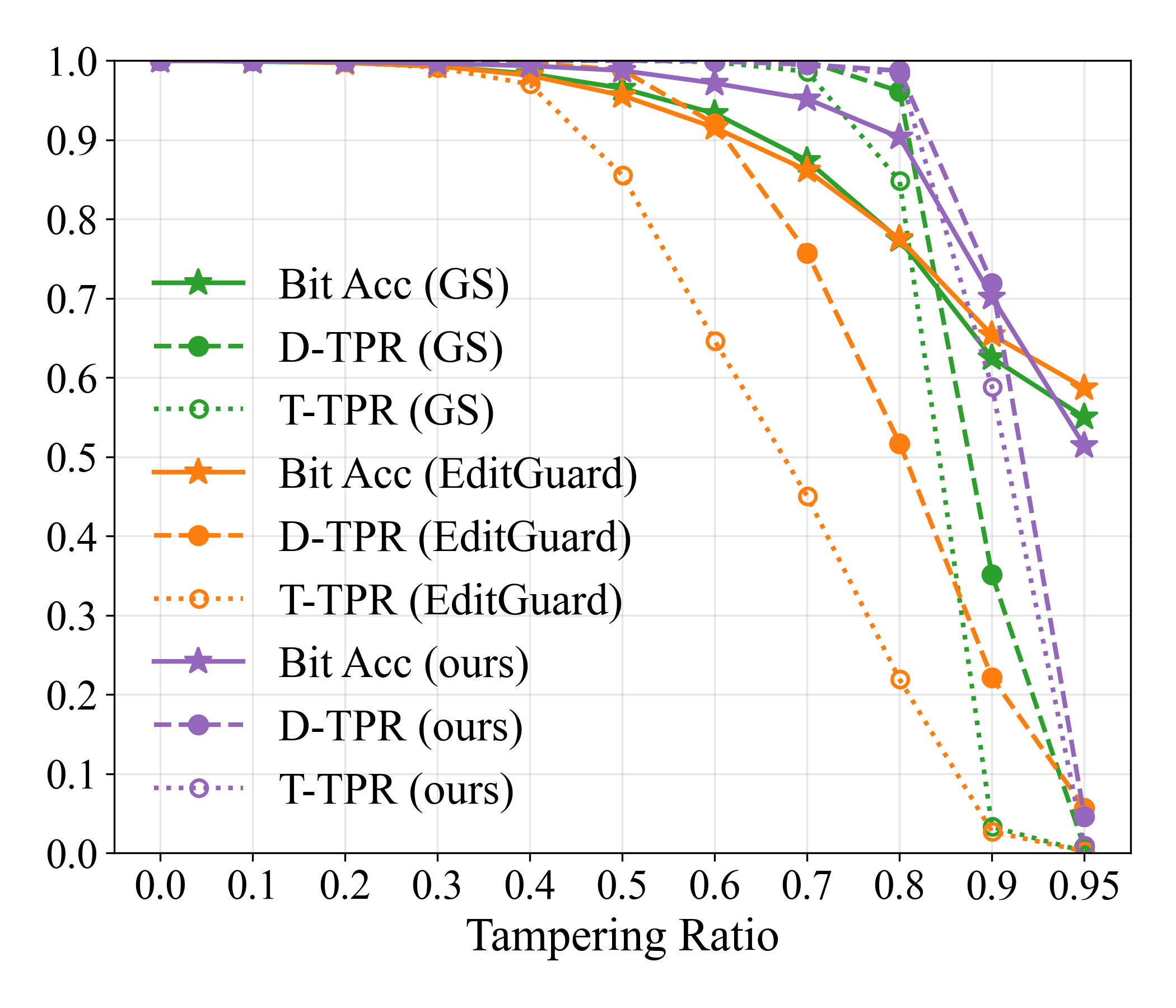}
        \caption{Foreground Tampering(Clean)}
        \label{fig:fg_clean}
    \end{subfigure}
    \hfill
    \begin{subfigure}[t]{0.246\textwidth}
        \includegraphics[width=\linewidth]{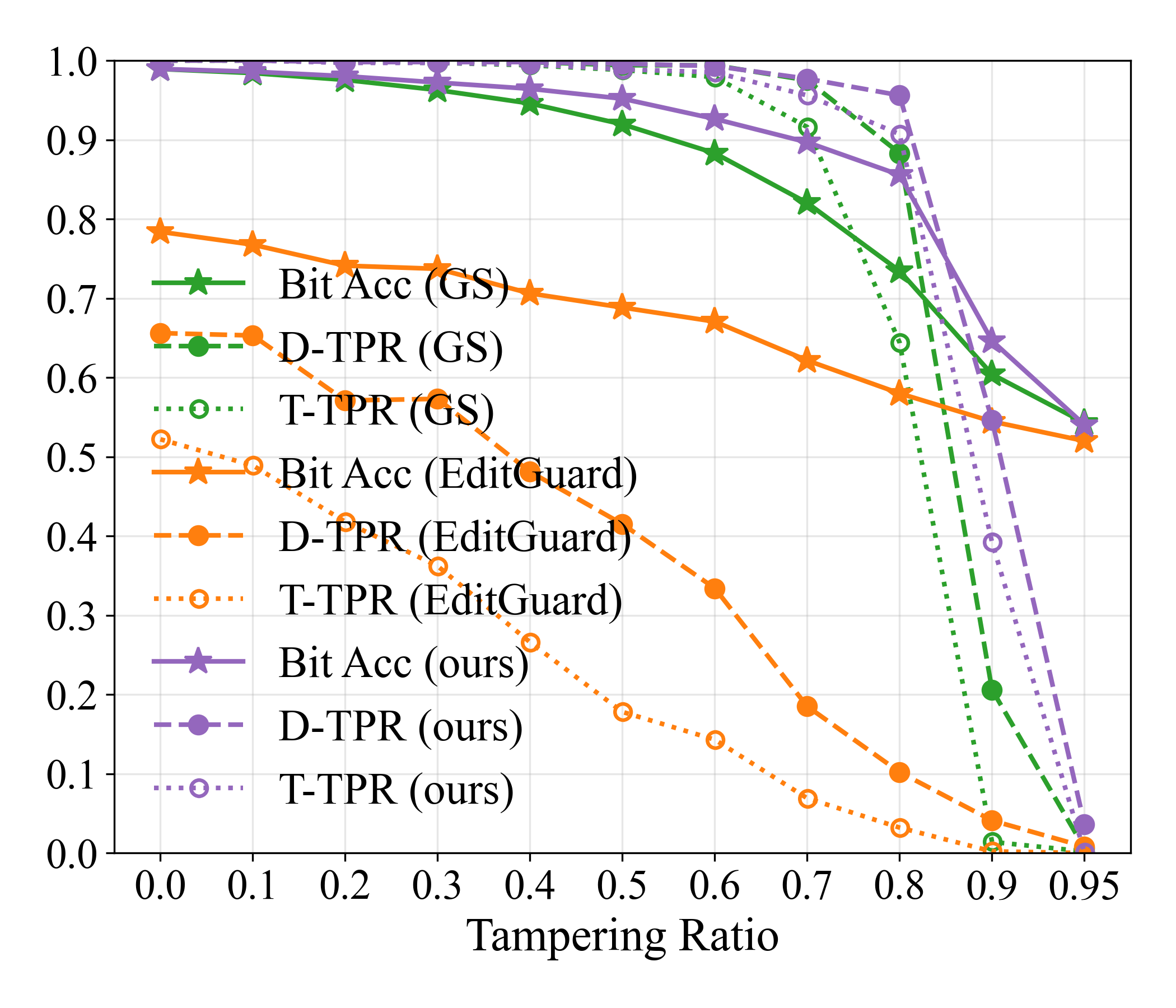}
        \caption{Foreground Tampering(Degraded)}
        \label{fig:fg_degraded}
    \end{subfigure}
    \hfill
    \begin{subfigure}[t]{0.246\textwidth}
        \includegraphics[width=\linewidth]{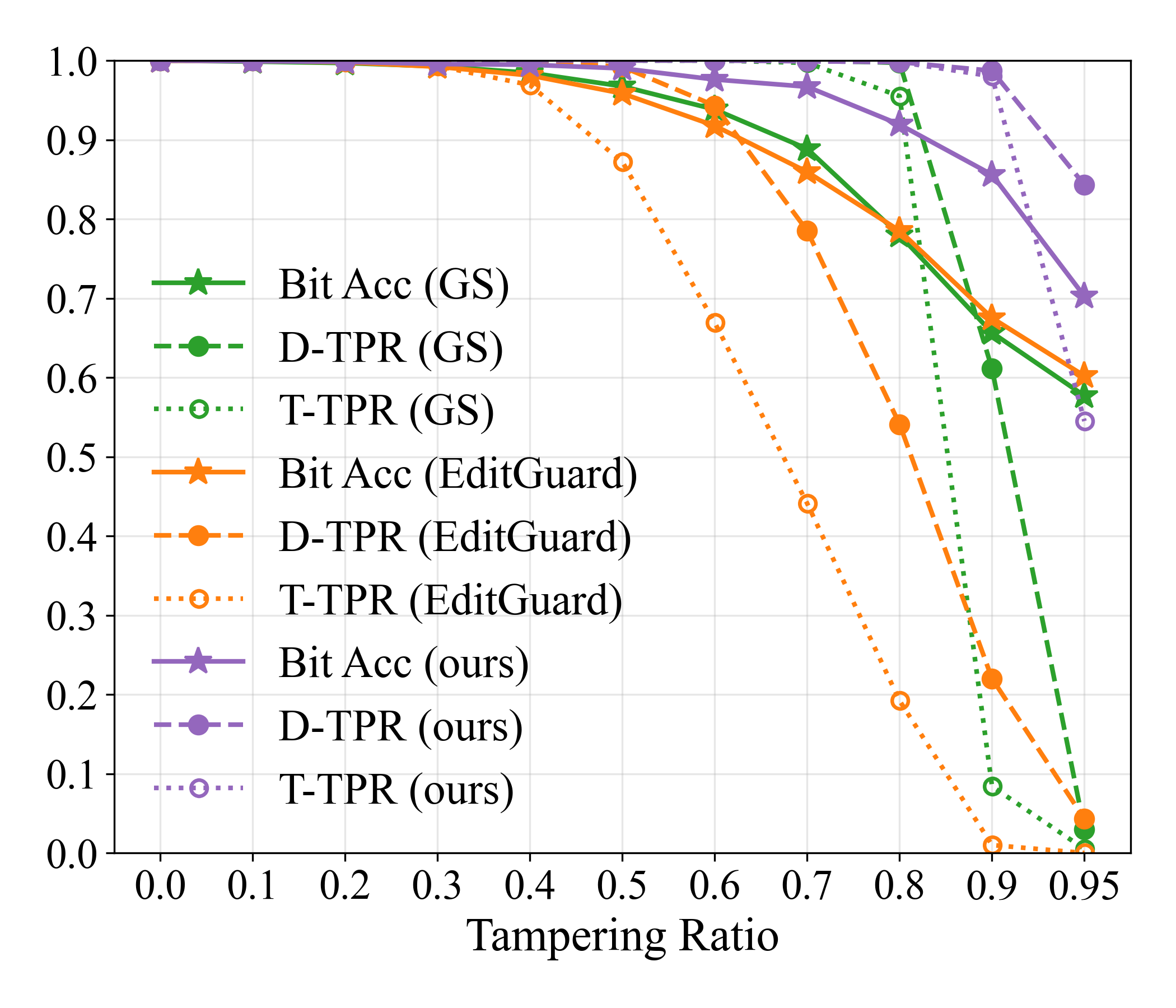}
        \caption{Background Tampering(Clean)}
        \label{fig:bg_clean}
    \end{subfigure}
    \hfill
    \begin{subfigure}[t]{0.247\textwidth}
        \includegraphics[width=\linewidth]{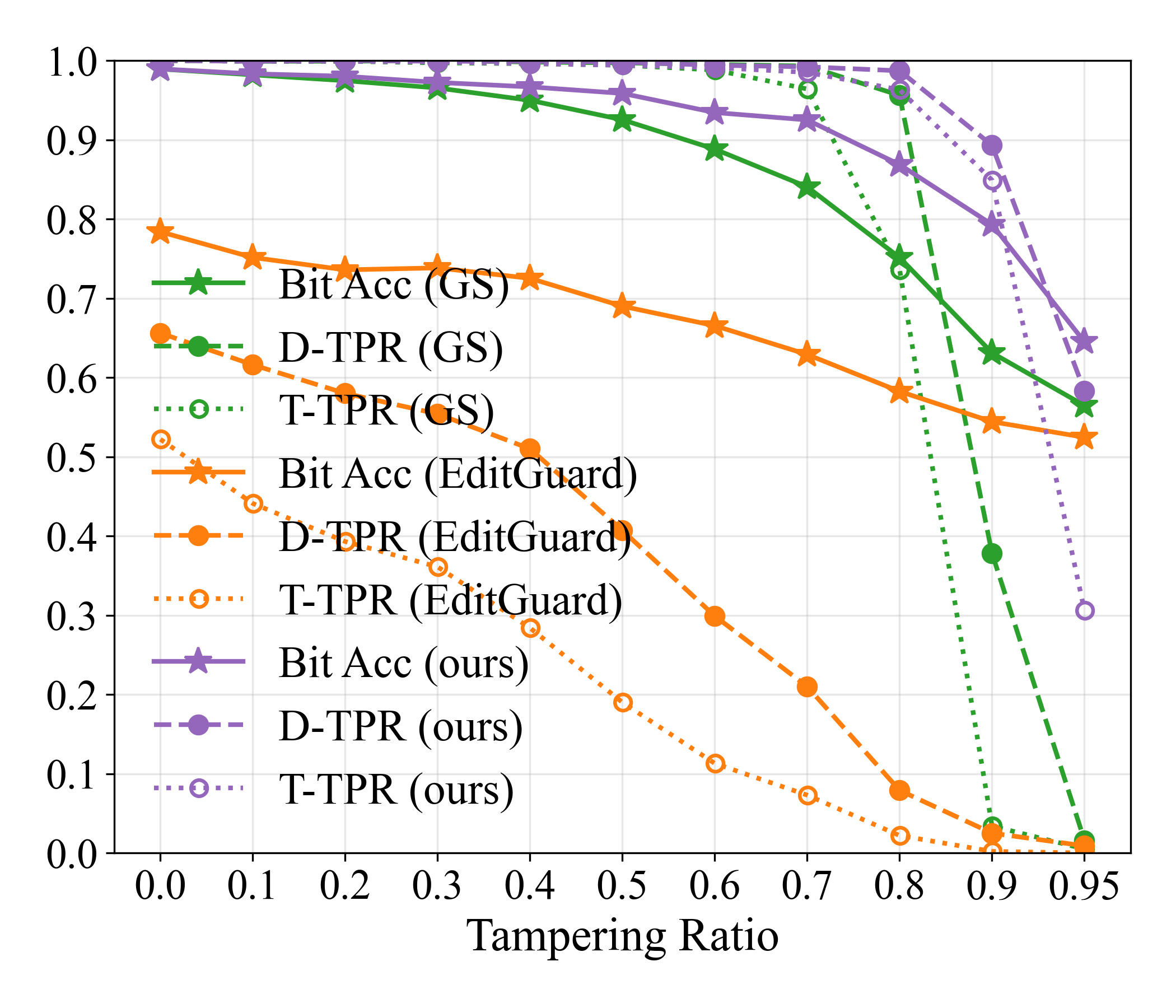}
        
        \caption{Background Tampering(Degraded)}
        \label{fig:bg_degraded}
    \end{subfigure}
    \vspace{-0.2cm}
    \caption{Comparative results in tampering scenarios using Gaussian Shading (GS) and EditGuard. We evaluate two types of tampering at ten different ratios for both clean and degraded images. The ``Tampering Ratio'' refers to the ratio of the area of tampering to the total image area.}
    \label{fig:Copyright_Watermarking_Resistance_to_ITAs}
    \vspace{-0.2cm}
\end{figure*}

\begin{figure*}[htbp]
    \centering
    \includegraphics[width=\linewidth]{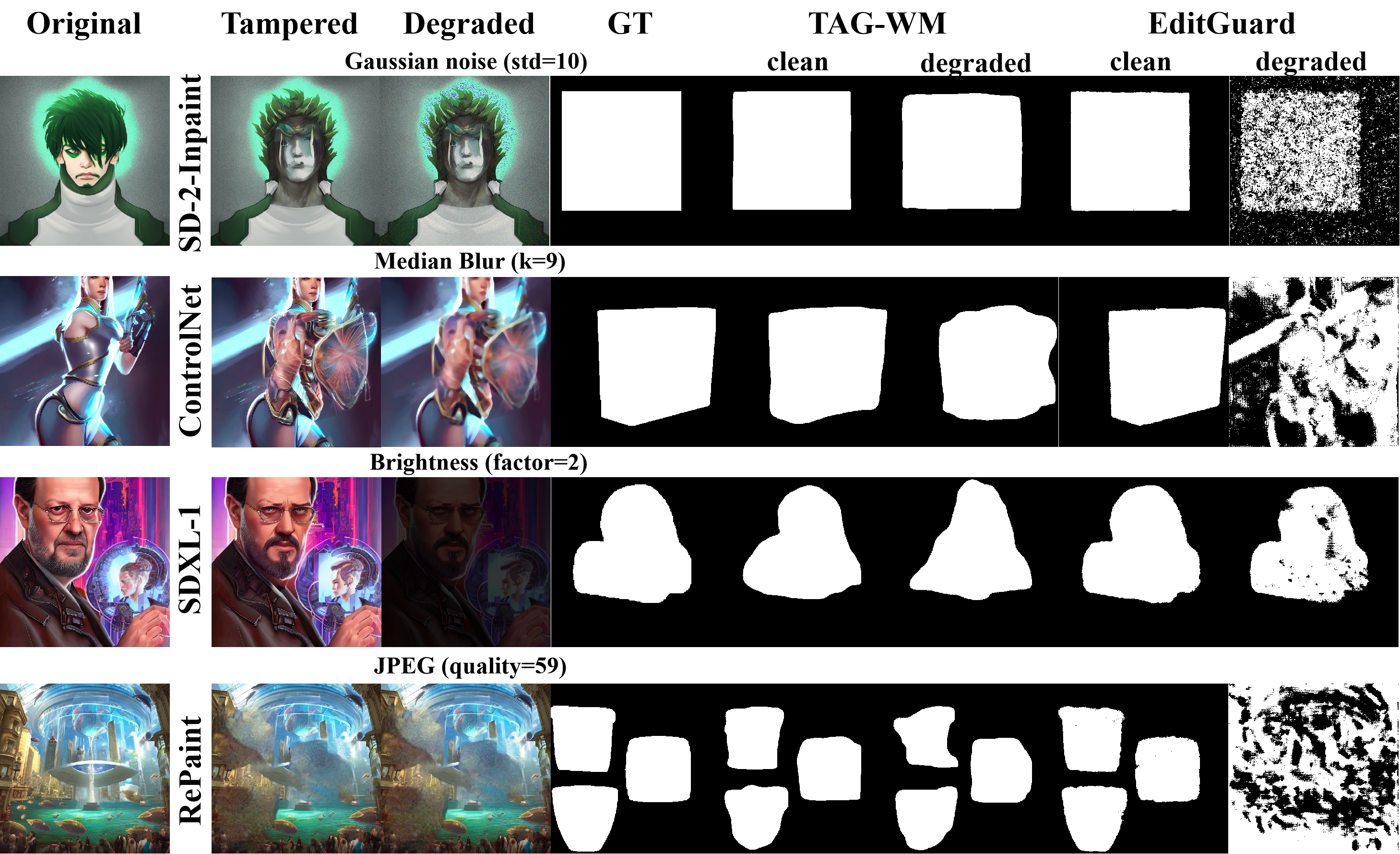}
    \caption{Visual comparison results with EditGuard. \textbf{Degraded} refers to images that have been tampered with and subsequently degraded, with the type and degree of degradation labeled at the top. The \textbf{clean} and \textbf{degraded} labels below the two methods refer to the predicted localization results for tampered images and degraded tampered images, respectively.}
    \label{fig: Visual comparative results with tampering localization methods}
\end{figure*}

\begin{table*}[htbp]
\centering
\resizebox{\textwidth}{!}{
\setlength{\tabcolsep}{.1in}{
\begin{tabular}{@{}clccccccccccccc@{}}
\toprule
\multicolumn{2}{c}{\multirow{2}{*}{\textbf{Methods}}} & \multicolumn{5}{c}{\textbf{Clean}} & \multicolumn{5}{c}{\textbf{Degraded}}   \\ \cmidrule(r){3-7} \cmidrule(l){8-12}
\multicolumn{2}{c}{}                                  &  \textbf{AUC} & \textbf{IoU} & \textbf{Dice} & \textbf{Bit Acc (\%)} & \textbf{T-TPR} & \textbf{AUC} & \textbf{IoU} & \textbf{Dice} & \textbf{Bit Acc (\%)} & \textbf{T-TPR} \\ 
\midrule
\multicolumn{12}{c}{\textbf{ControlNet-v11p-sd15-Inpaint}} \\ \midrule
\multicolumn{2}{c}{EditGuard} &
\textbf{0.9755} & \textbf{0.9854} & \textbf{0.9925} & 82.639 & \multicolumn{1}{c|}{0.477} & 
0.6152 &0.7620 &0.8425 &62.561 &0.129 
\\
\multicolumn{2}{c}{TAG-WM (ours)} &
0.9697 & 0.9727 & 0.9858 & \textbf{92.327} &  \multicolumn{1}{c|}{\textbf{0.989}} &  \textbf{0.9118} & \textbf{0.9374} & \textbf{0.9650} & \textbf{86.726} & \textbf{0.907} \\
\midrule
\multicolumn{12}{c}{\textbf{Stable Diffusion-2-Inpainting}} \\ \midrule
\multicolumn{2}{c}{EditGuard} &
\textbf{0.9772} &\textbf{0.9843} &\textbf{0.9918} &82.672 & \multicolumn{1}{c|}{0.485} &
0.6187 &0.7665 &0.8461 &63.583 &0.139 
\\
\multicolumn{2}{c}{TAG-WM (ours)}  &
0.9759&0.9776&0.9882&\textbf{92.621}&\multicolumn{1}{c|}{\textbf{0.990}} &\textbf{0.9204} &\textbf{0.9426} &\textbf{0.9682} &\textbf{87.065} &\textbf{0.914} \\
\midrule
\multicolumn{12}{c}{\textbf{SDXL-1-Inpainting}} \\ \midrule
\multicolumn{2}{c}{EditGuard} &  
\textbf{0.9767} &\textbf{0.9847} &\textbf{0.9920} &82.769 &\multicolumn{1}{c|}{0.477} &
0.6116 &0.7596 &0.8406 &62.867 &0.131 \\
\multicolumn{2}{c}{TAG-WM (ours)}  &
0.9753 &0.9770 &0.9880 & \textbf{92.986} &\multicolumn{1}{c|}{\textbf{0.991}}  &\textbf{0.9174} &\textbf{0.9407} &\textbf{0.9669} &\textbf{87.288} &\textbf{0.913} 
\\
\midrule
\multicolumn{12}{c}{\textbf{RePaint}} \\ \midrule
\multicolumn{2}{c}{EditGuard} & 
\textbf{0.9636} &\textbf{0.9839} &\textbf{0.9917} & 80.461 & \multicolumn{1}{c|}{0.428} &
0.6179 &0.7677 &0.8461 &63.064 &0.133 \\
\multicolumn{2}{c}{TAG-WM (ours)}  &
0.9723 &0.9765 & 0.9878 &\textbf{91.638} & \multicolumn{1}{c|}{\textbf{0.987}} &\textbf{0.9154} &\textbf{0.9401} &\textbf{0.9669} &\textbf{86.305} &\textbf{0.910}
\\
\bottomrule
\end{tabular}
}}
\caption{Comparative results with EditGuard for both clean and degraded scenarios (all tampering types are zero-shot).}
\label{table: Comparative results with EditGuard}
\end{table*}

\noindent\textbf{Baseline Methods}.
We select seven baseline methods: three officially used by SD, namely DwtDct \cite{old_wm-digital_wm_and_stega} and DwtDctSvd \cite{old_wm-digital_wm_and_stega}; two post-processing-based methods RivaGAN \cite{postprocessing_watermarking-RivaGAN} and EditGuard \cite{wm_and_tl-Editguard}, a fine tuning-based method called Stable Signature \cite{fine_tuning_based_watermarking-StableSignature}, a latent representation-based method called Tree-Ring \cite{latent_representation_based_watermarking-Treering}, and GS \cite{gaussianshading}.

\noindent\textbf{Evaluation Metrics}. 
To measure the performance of watermarking methods, we calculated the bit accuracy (Bit Acc). To measure the bias in model performance, we computed the CLIP-Score \cite{CLIP} for 10 batches of watermarked images and performed a t-test on the mean CLIP-Score compared to that of watermark-free images. In prior works, the incorporation of watermark embedding
modules inevitably results in a decline in model performance, therefore typically evaluated using the Peak Signal-to-Noise Ratio (PSNR) and Fréchet Inception Distance (FID) \cite{metrics-FID}, which are unnecessary to our method. To simulate detecting and tracing scenarios, we also calculated the detecting true positive rate (D-TPR) and the tracing true positive rate (T-TPR), by fixing the false positive rate (FPR) and the number of users at $10^{-6}$ and $10^{6}$, separately \cite{gaussianshading}. 

\noindent\textbf{Performance in Non-Tampering Scenarios.}  
Table \ref{table: Overall comparative results with watermarking methods} demonstrates the effectiveness of our method compared to various copyright watermarking techniques under both clean and degraded conditions, with no tampering involved. 

\noindent\textbf{Performance in Tampering Scenarios.}  
Furthermore, we compare the performance of TAG-WM in tampering scenarios with state-of-the-art methods, including the diffusion inversion-based GS and EditGuard, a method specifically designed for these scenarios. Fig. \ref{fig:Copyright_Watermarking_Resistance_to_ITAs} illustrates the results.

\underline{\textbf{(1) Across all four scenarios}}, our method consistently outperforms others. As the tampering ratio increases to 0.7, TAG-WM surpasses GS in Bit Accuracy by over 7\%, and this advantage grows to 10\% when the ratio reaches 0.8. EditGuard maintains a similar bit accuracy to GS in clean scenarios but performs significantly worse in degraded scenarios. Additionally, due to limitations in bit capacity, even with comparable bit accuracy, its D-TPR and T-TPR are much lower than GS and our method.
\underline{\textbf{(2) In clean scenarios}}, when the tampering ratio rises to 0.7, GS's Bit Accuracy falls below 90\%, while TAG-WM remains above 95\%, demonstrating competitive performance.  
\underline{\textbf{(3) In clean background tampering scenarios}}, TAG-WM achieves 0.987 in D-TPR and 0.980 in T-TPR, significantly outperforming GS (0.611 in D-TPR, 0.084 in T-TPR). \underline{\textbf{(4) Even under degraded background tampering}}, TAG-WM maintains strong resistance, achieving 0.893 in D-TPR and 0.849 in T-TPR, again vastly exceeding GS (0.378 in D-TPR, 0.034 in T-TPR). These results highlight TAG-WM's robustness and adaptability to extreme conditions.  
\underline{\textbf{(5) Interestingly, in foreground tampering scenarios}}, TAG-WM's performance degrades more than in background tampering when the tampering ratio is high. Given our TAD algorithm, we attribute this to the weaker tampering localization ability in the former. The performance gap likely stems from differences in training data: we collected more small tampering masks for foreground tampering, limiting the ability to detect large tampered areas, whereas for background tampering, we collected more large masks, leading to stronger performance.

\begin{table*}[htbp]
\centering
\resizebox{\textwidth}{!}{
\setlength{\tabcolsep}{.04in}{
\begin{tabular}{@{}cccccccccccccccccc@{}}
\toprule
\multicolumn{2}{c}{\multirow{2}{*}{\textbf{Settings}}} & \multicolumn{8}{c}{\textbf{Trainable}} & \multicolumn{8}{c}{\textbf{Train-free}}   \\ \cmidrule(r){3-10} \cmidrule(l){11-18}
\multicolumn{2}{c}{}                                  & \textbf{Acc} & \textbf{Pre} & \textbf{Spe} & \textbf{Rec} & \textbf{AUC} & \textbf{IoU} & \textbf{Dice} & \textbf{Average} & \textbf{Acc} & \textbf{Pre} & \textbf{Spe} & \textbf{Rec} & \textbf{AUC} & \textbf{IoU} & \textbf{Dice} & \textbf{Average} \\ 
\midrule
\multicolumn{18}{c}{\textbf{Crop}} \\ \midrule
\multicolumn{2}{c}{3 intervals} 
& 0.9964& 0.9988& \textbf{0.9568} & \textbf{0.9934}& \textbf{0.9751} & 0.9923 & 0.9961 & \multicolumn{1}{c|}{\textbf{0.98699}} &
\textbf{0.9146}&\textbf{0.8935}&\textbf{0.6522}&\textbf{0.9200}&\textbf{0.7861}&\textbf{0.8352}&\textbf{0.9014}&\textbf{0.84329}\\
\multicolumn{2}{c}{4 intervals}  
&\textbf{0.9965}&\textbf{0.9991}&0.9561&0.9933&0.9747&\textbf{0.9924}&\textbf{0.9962}&\multicolumn{1}{c|}{0.98690}&0.8627&0.8529&0.6335&0.8755&0.7545&0.7677&0.8566&0.80049 \\
\midrule
\multicolumn{18}{c}{\textbf{Drop}} \\ \midrule
\multicolumn{2}{c}{3 intervals}  
& \textbf{0.9972} & \textbf{0.9961} & 0.9987 & \textbf{0.9786} & \textbf{0.9887} & \textbf{0.9765} & \textbf{0.9871} & \multicolumn{1}{c|}{\textbf{0.98899}} & 
\textbf{0.9039}&\textbf{0.7221}&\textbf{0.8301}&\textbf{0.8257}&\textbf{0.8279}&\textbf{0.6328}&\textbf{0.7356}&\textbf{0.78259}\\
\multicolumn{2}{c}{4 intervals} 
&0.9966&0.9786&\textbf{0.9989}&0.9602&0.9795&0.9581&0.9689&\multicolumn{1}{c|}{0.97726}&0.8491&0.5975&0.7780&0.7991&0.7886&0.5286&0.6418&0.71181 
\\ 
\midrule
\multicolumn{18}{c}{\textbf{Logo Insertion}} \\ \midrule
\multicolumn{2}{c}{3 intervals}  
& \textbf{0.9849} & \textbf{0.9776} & \textbf{0.9950} & \textbf{0.9268} & \textbf{0.9609} & \textbf{0.9101} & \textbf{0.9506} & \multicolumn{1}{c|}{\textbf{0.95799}} & \textbf{0.8515}&\textbf{0.6319}&\textbf{0.8400}&\textbf{0.8352}&\textbf{0.8376}&\textbf{0.5632}&\textbf{0.7019}&\textbf{0.75161}\\
\multicolumn{2}{c}{4 intervals}
&0.9819&0.9730&0.9945&0.9076&0.9511&0.8904&0.9374&\multicolumn{1}{c|}{0.94799}&0.7806&0.5234&0.7588&0.8009&0.7798&0.4638&0.6130&0.67433 \\
\midrule
\multicolumn{18}{c}{\textbf{MAT}} \\ \midrule
\multicolumn{2}{c}{3 intervals}  
& \textbf{0.9897} & \textbf{0.9946} & \textbf{0.9964} & \textbf{0.9755} & \textbf{0.9860} & \textbf{0.9705} & \textbf{0.9848} & \multicolumn{1}{c|}{\textbf{0.98536}} & 
\textbf{0.8573}&\textbf{0.8056}&\textbf{0.8453}&\textbf{0.8171}&\textbf{0.8312}&\textbf{0.6782}&\textbf{0.7987}&\textbf{0.80477} \\
\multicolumn{2}{c}{4 intervals}
&0.9868&0.9911&0.9943&0.9700&0.9821&0.9621&0.9803&\multicolumn{1}{c|}{0.98096}&0.7953&0.7085&0.7675&0.7807&0.7741&0.5892&0.7293&0.73494\\
\bottomrule
\end{tabular}
}}
\caption{Performance of the tampering localization with different sampling and DVRD settings.
}
\label{table: Comparison Results in the Number of Embedded Intervals of the Tampering Localization Template}
\end{table*}

\subsection{Localization Performance}
\label{Localization Performance}

\noindent\textbf{Baseline Methods}. 
Previous research \cite{wm_and_tl-Editguard} has shown that passive methods fail to generalize to unseen tampering types that differ from those encountered during training. For state-of-the-art (SOTA) passive tampering localization methods, such as MVSS-Net \cite{tl-passive-universal-MVSSNet}, OSN \cite{tl-passive-specific-OSN}, PSCC-Net \cite{tl-passive-universal-PSCCNet}, and HiFi-Net \cite{tl-passive-universal-HiFiNet}, the Intersection over Union (IoU) is consistently below 0.3 when tested on unseen optimal image inpainting methods, including ControlNet \cite{ImageInpainting-ControlNet}, Stable Diffusion Inpainting \cite{DMs-LDMs}, and SDXL \cite{ImageInpainting-SDXL}. Similarly, the Dice score remains below 0.65. Therefore, as a proactive method, we compare our approach exclusively with the SOTA proactive tampering localization method, EditGuard \cite{wm_and_tl-Editguard}, where all tampering types remain unseen during training, with tampering ratios ranging from 0.3 to 0.7.

\noindent\textbf{Evaluation Metrics}. 
To evaluate the performance of tampering localization, we calculated Area Under the Curve (AUC), Intersection over Union (IoU), and Dice score (Dice). To assess the performance of copyright message extraction and the robustness of the methods to image degradations, we randomly applied a series of image degradations with varying degrees. Since EditGuard is not only a SOTA proactive tampering localization method but also a dual-watermark framework similar to ours, we include additional comparisons by reporting Bit Acc and T-TPR for copyright message recovery.

The statistical comparative results are shown in Table \ref{table: Comparative results with EditGuard}, demonstrating that our method exhibits comparable zero-shot capability to EditGuard under various conditions. \underline{\textbf{(1) In clean scenarios}}, both methods show excellent localization performance, with Dice scores exceeding 98.5\%. Our method achieves a score just slightly lower than EditGuard, by less than 1\%, while outperforming EditGuard by approximately 10\% in Bit Accuracy, indicating a more robust ability to preserve copyright messages. \underline{\textbf{(2) In degraded scenarios}}, our method, TAG-WM, consistently outperforms EditGuard in both tampering localization and copyright message extraction. This is an interesting observation, as our method is trained solely on clean data yet demonstrates better generalization to degraded data. This advantage arises from our unique design, which includes utilizing diffusion inversion sensitivity, mapping tampered data from image space to mask space, and decoupling image distribution from tampering localization. These strategies contribute to the natural, train-free generalization observed in our method. 

Fig.\ref{fig: Visual comparative results with tampering localization methods} visually presents the predicted results of the two methods under the aforementioned tampering types and specific image quality degradations. It further illustrates that in clean scenarios, TAG-WM exhibits only slight errors compared to EditGuard. However, in most degradation scenarios, TAG-WM maintains its effectiveness, while EditGuard almost loses its capability. 
Detailed robustness evaluation under different image degradations is illustrated in Appendix \hyperref[sup: Detailed Robustness Evaluation]{B}.

\subsection{Training Details for Trainable DVRD}
\label{Training Details for Trainable DVRD}
\paragraph{Dataset}
\label{Dataset}

Firstly, we constructed the training dataset using 5,000 diverse prompts from the Stable-Diffusion-Prompt dataset. Each prompt was used to generate an image with the Stable Diffusion v2.1 model under five different ODE schedulers—DDIM \cite{DMs-scheduler-DDIM}, UniPC \cite{DMs-scheduler-UniPC}, PNDM \cite{DMs-scheduler-PNDM}, DEIS \cite{DMs-scheduler-DEIS}, and DPMSolver \cite{DMs-scheduler-DPMSolver}—resulting in a total of 5,000 images (1,000 images per scheduler). 
Secondly, Each generated image underwent all of the following tampering types: random cropping (\textit{crop ratio=0.1, 0.3, 0.5, 0.7 and 0.9}), random pixel dropping (\textit{drop ratio=0.1, 0.3, 0.5, 0.7 and 0.9}), random logo insertion with varying numbers and sizes (\textit{logo count-logo ratio=1-0.7, 3-0.39, 5-0.25, 7-0.2 and 9-0.1}), and image inpainting using MAT \cite{ImageInpainting-MAT} pre-trained on three datasets—CelebA-HQ\footnote{https://www.kaggle.com/datasets/vincenttamml/celebamaskhq512}, FFHQ\footnote{https://huggingface.co/datasets/LIAGM/FFHQ\_datasets}, and Places365-Standard\footnote{https://paperswithcode.com/dataset/places365}—all at a resolution of $512\times 512$. These tampering techniques simulate a variety of altered regions with different shapes, sizes, and edge smoothness. Note that our proactive tampering localization method does not require distinguishing between tampering types; thus, the aforementioned manipulations are sufficient for our purposes. For each tampered image, we also stored the corresponding ground truth tampering mask. Finally, we split the dataset into training and validation sets at a ratio of $0.95:0.05$.

\noindent\textbf{Training Settings}. 
We conducted the training on an NVIDIA GeForce RTX 2080 Ti. We use mean squared error as the loss function and fix the learning rate to $10^{-3}$. Both the input data and labels were resized to a fixed size of 64 before being fed into the network. The batch size was set to 256, and the model was trained for a total of 500 epochs.


\subsection{Impact of Sampling and DVRD Strategy} 
\label{Impact of Training and Number of Intervals}
We evaluated the impact of the number of embedded intervals of $W_{loc}$ for both the train-free DVRD and the trainable DVRD in the validation set. Table \ref{table: Comparison Results in the Number of Embedded Intervals of the Tampering Localization Template} clearly shows that the trainable DVRDs utilizing three-intervals strategy consistently outperform other counterparts, which proves our analyses and illustrates the superiority of our strategies.


\section{Conclusion}
\label{Conclusion}
In this paper, we present the first in-generation image watermarking framework that integrates copyright message embedding and tamper localization within diffusion models. Our approach achieves several notable advancements: first, we introduce parallel watermark embedding, which eliminates the mutual interference between the two watermarks and provides a 4$\times$ increase in capacity compared to post-processing methods like EditGuard, without any loss in image quality. Second, we propose a tampering-aware optimization strategy, which dynamically adjusts watermark robustness based on tampering localization, resulting in a 7-10\% improvement in tamper resistance across various tampering scenarios compared to state-of-the-art methods. Finally, our framework is highly efficient, operating at just 10.78ms per image during generation, which is 6.3$\times$ faster than the post-processing approach EditGuard.
However, the reliance on deterministic DDIM inversion limits compatibility with other SDE schedulers, and like existing in-generation methods, our framework remains vulnerable to full-image adversarial attacks that globally distort noise patterns. And using a fixed seed to initialize localization watermarks may make it vulnerable to malicious attacks. Future work involves enhancing fine-grained tampering detection through attention-guided watermark allocation and developing adaptive dynamic watermark strategies.



\section*{Appendix}
\subsection*{A. Implementation of Random Image Degradations}
\label{sup: Implementation of Random Image Degradations}
In the main experiments, we use random degradations to evaluate the performance of our method and baseline methods in degradation scenarios. The degradation type and strength are randomly selected from:
(1) Jpeg compression: $quality\in[30,90]$
(2) Gaussian noise: $mean=0, standard\in[1,5]$
(3) Gaussian blur: $radius\in\{1,2\}$
(4) Median blur: $kernel~size\in\{3, 5, 7, 9\}$
(5) Resize then recover: $ratio\in[0.6,0.9]$
(6) Brightness transformation: $factor\in\{1,2\}$. Only one degradation of one strength is applied to each sample, not a combination.

\subsection*{B. Detailed Robustness Evaluation}
\label{sup: Detailed Robustness Evaluation}
In this section, we validate the robustness of our method to image degradations in each type and specific strength. 

\begin{figure}[t!]
    \centering
    \begin{subfigure}[t]{0.23\textwidth}
        \includegraphics[width=\linewidth]{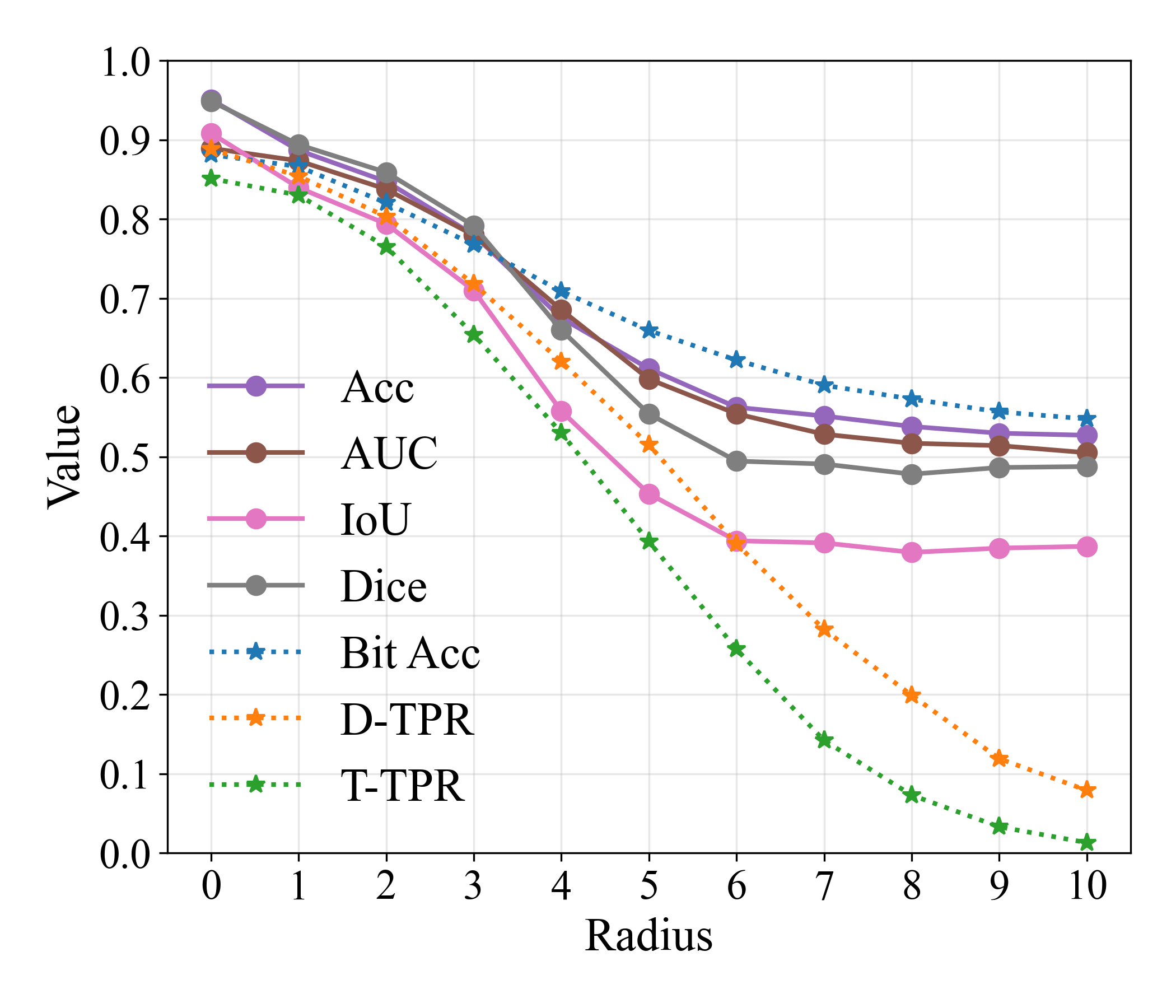}
        \caption{Gaussian Blur}
        \label{fig:sub1}
    \end{subfigure}
    \begin{subfigure}[t]{0.23\textwidth}
        \includegraphics[width=\linewidth]{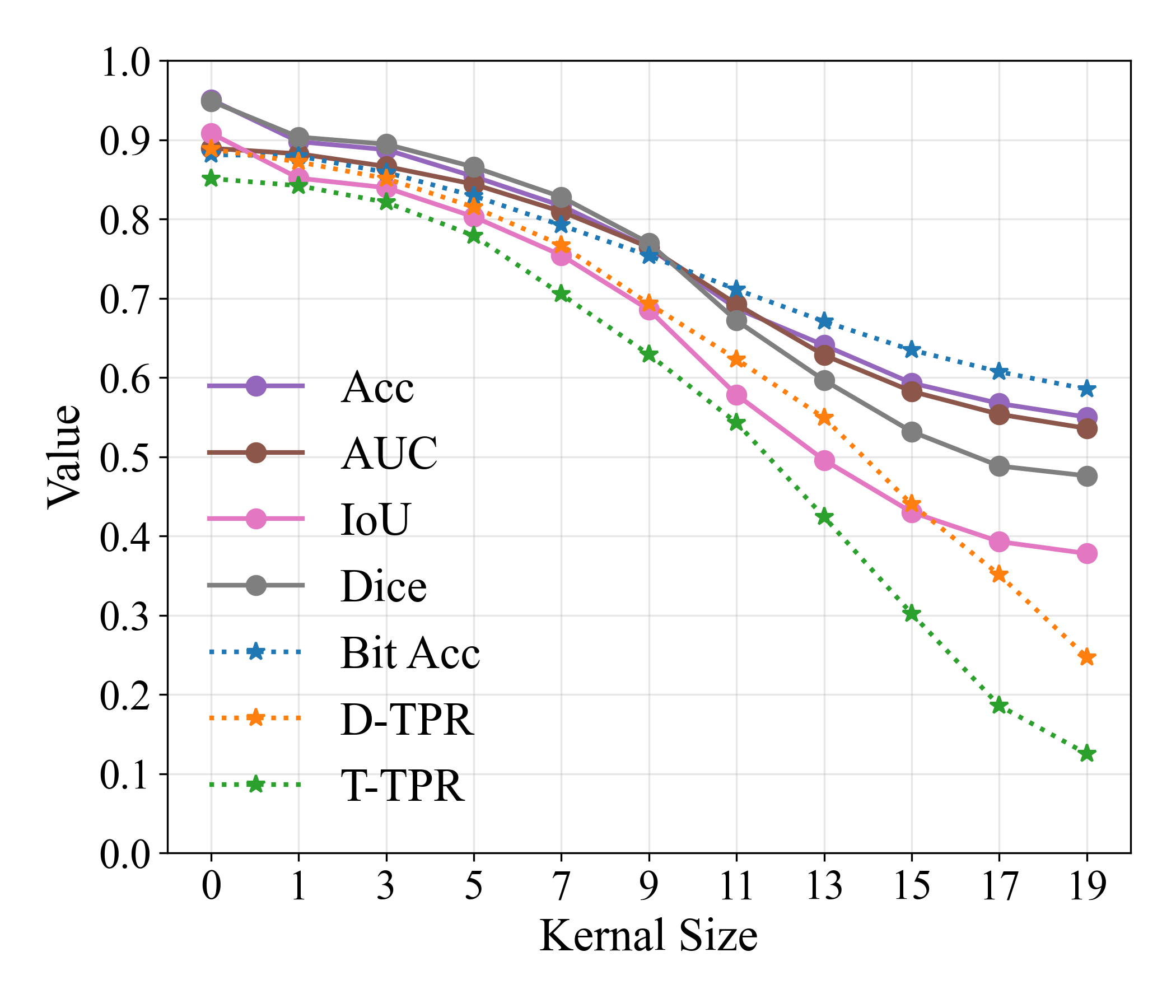}
        \caption{Median Filter}
        \label{fig:sub2}
    \end{subfigure}
    \\
    \begin{subfigure}[t]{0.23\textwidth}
        \includegraphics[width=\linewidth]{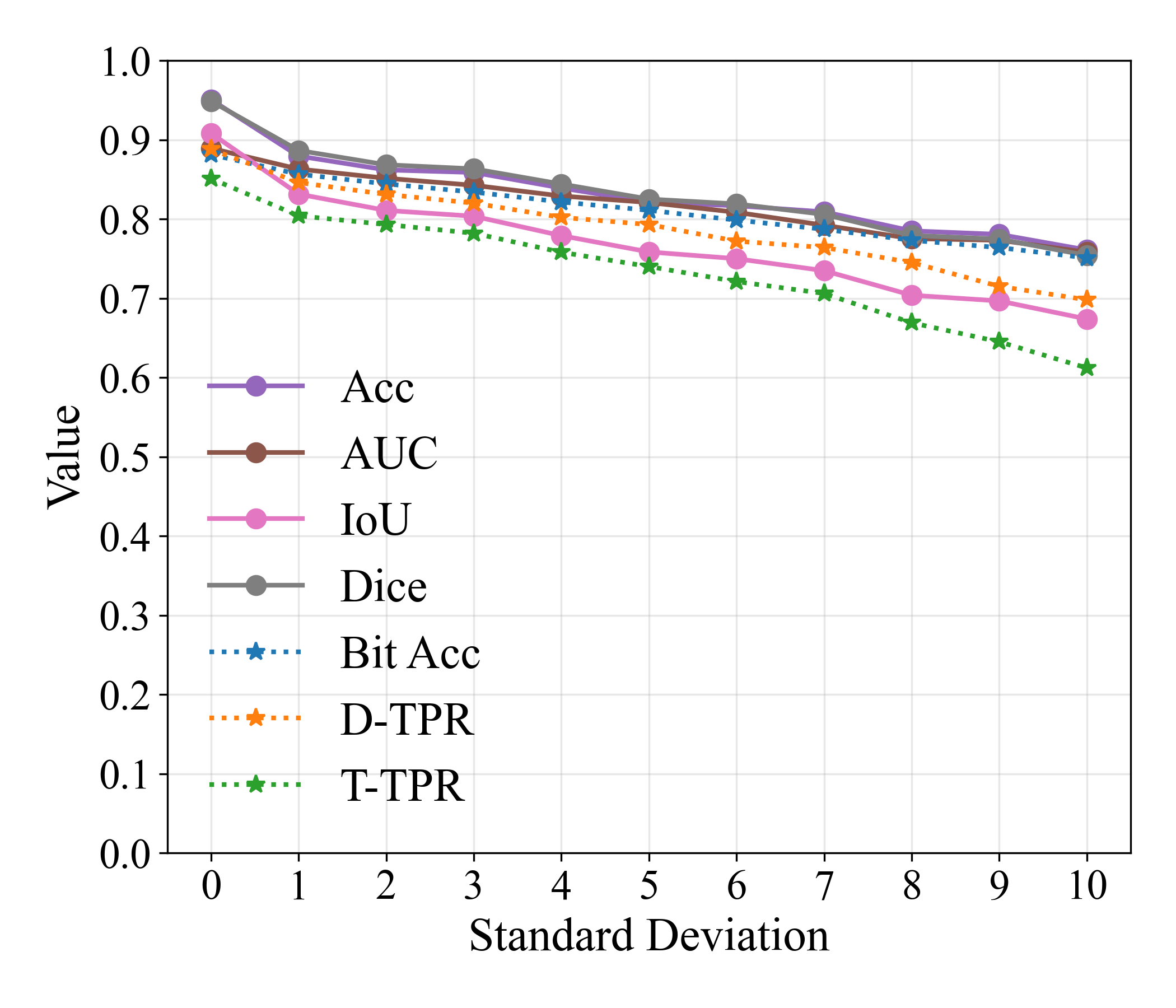}
        \caption{Gaussian Noise}
        \label{fig:sub3}
    \end{subfigure}
    \begin{subfigure}[t]{0.23\textwidth}
        \includegraphics[width=\linewidth]{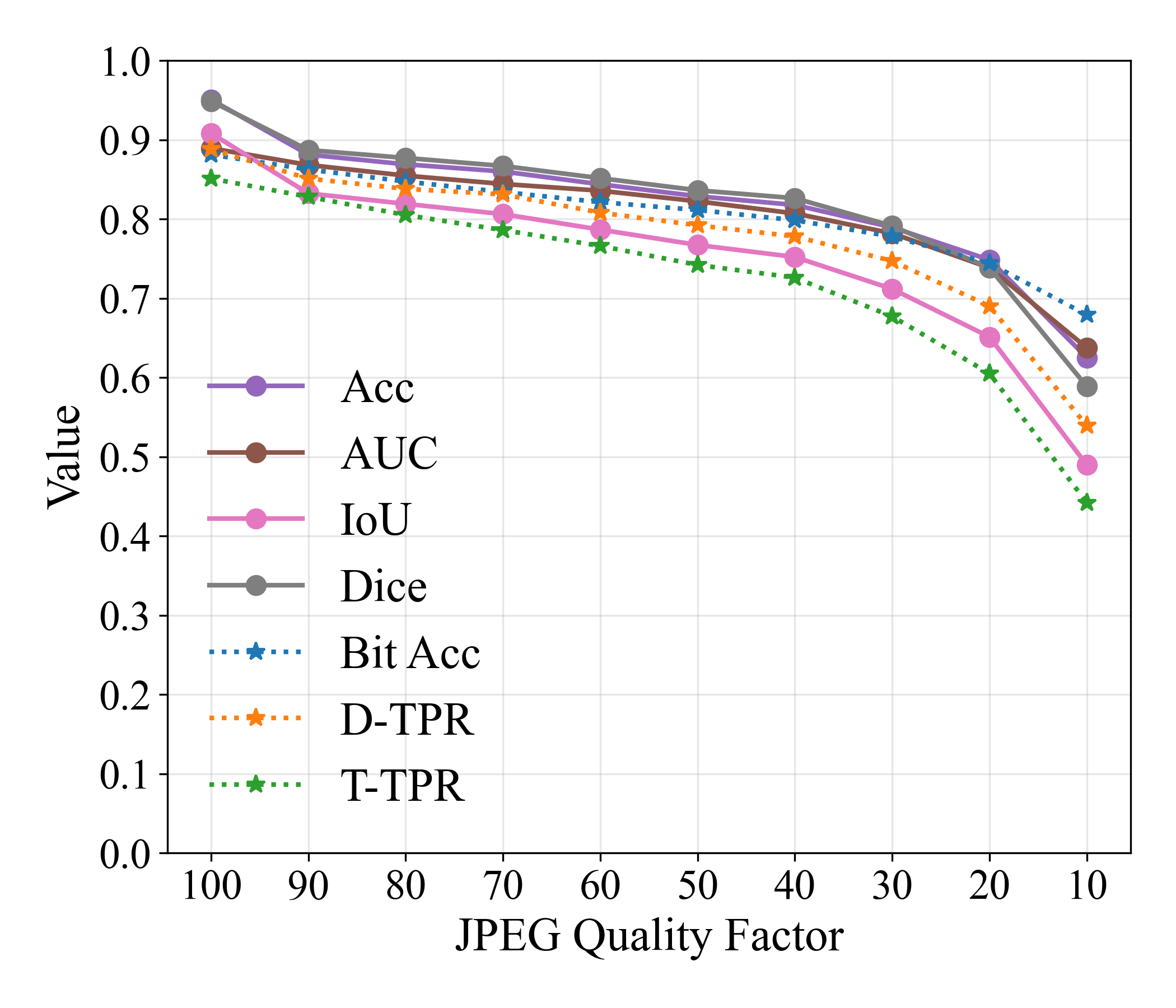}
        \caption{JPEG Compression}
        \label{fig:sub4}
    \end{subfigure}
    \\
    \begin{subfigure}[t]{0.23\textwidth}
        \includegraphics[width=\linewidth]{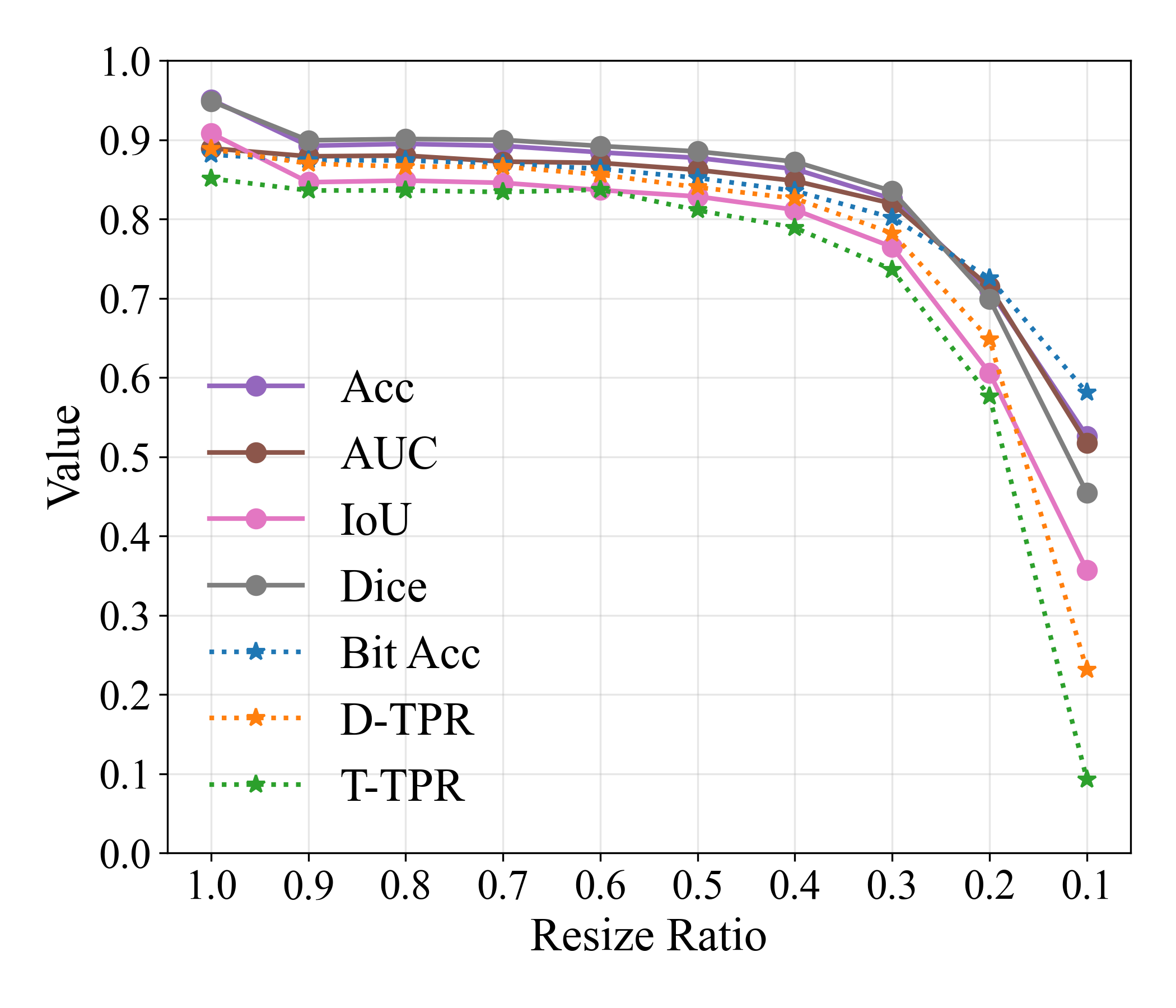}
        \caption{Resize then recover}
        \label{fig:sub5}
    \end{subfigure}
    \begin{subfigure}[t]{0.23\textwidth}
        \includegraphics[width=\linewidth]{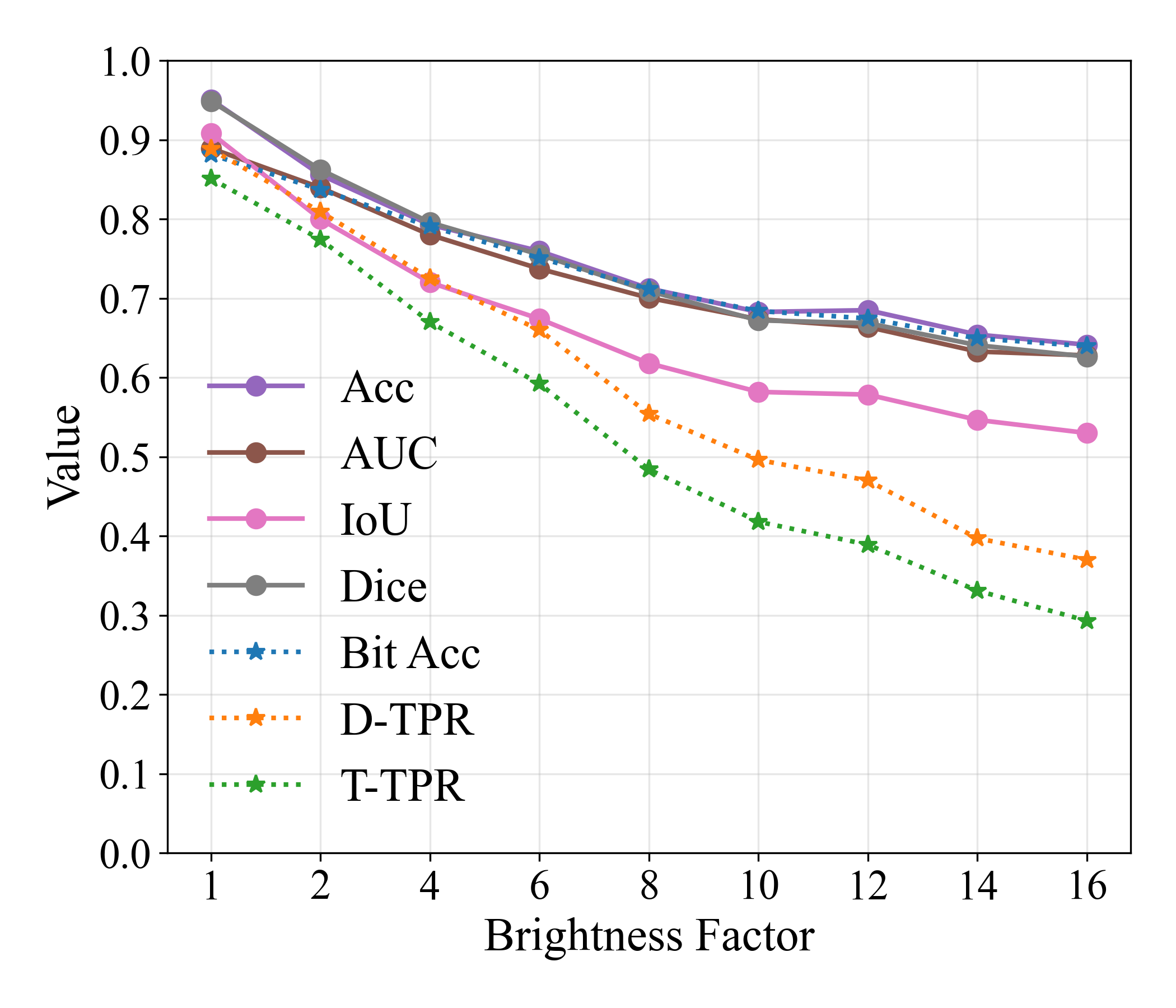}
        \caption{Brightness Transformation}
        \label{fig:sub6}
    \end{subfigure}
    \\
    \begin{subfigure}[t]{0.23\textwidth}
        \includegraphics[width=\linewidth]{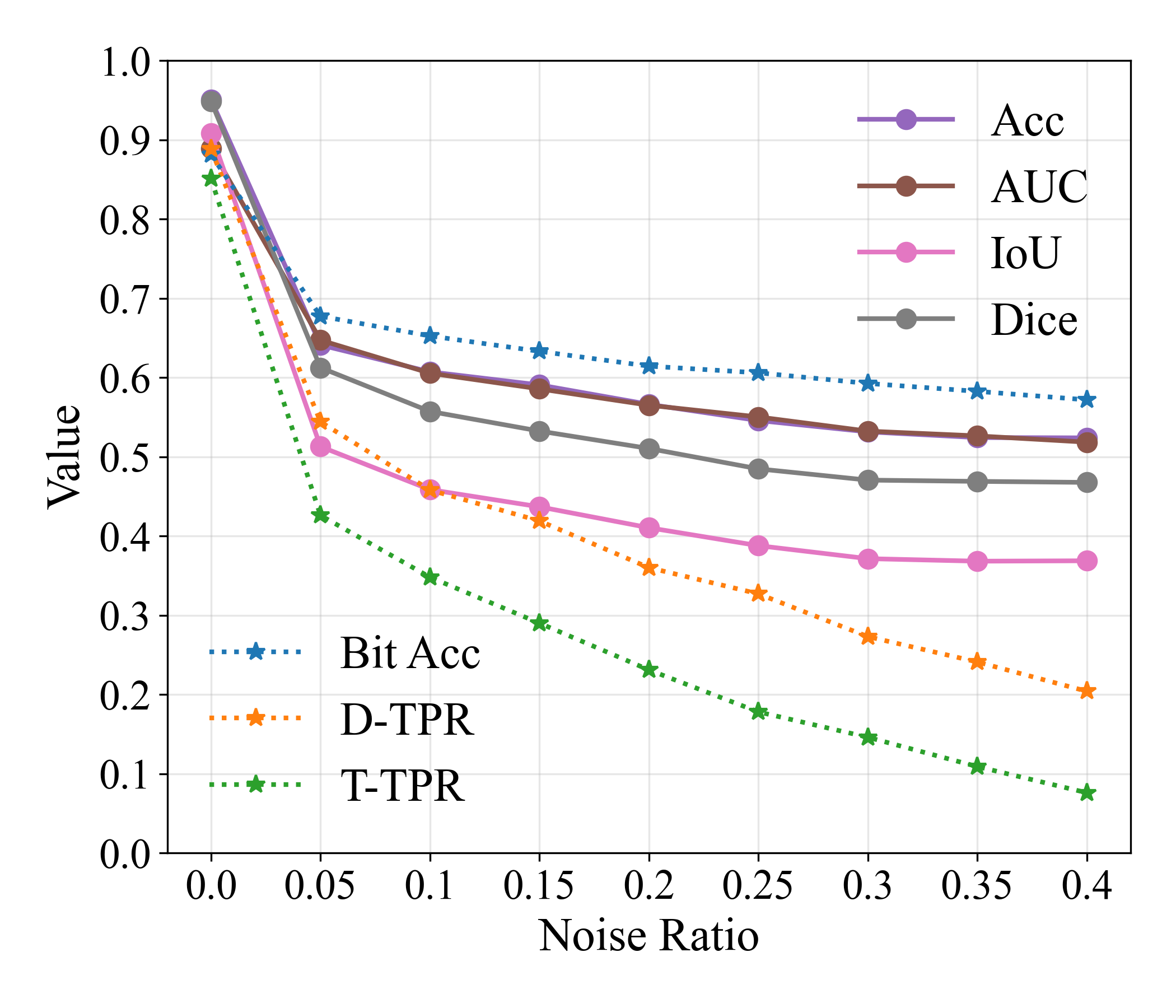}
        \caption{Salt and Pepper Noise}
        \label{fig:sub7}
    \end{subfigure}
    \caption{Robustness of our method to image degradations.}
    \label{fig: Robustness of our method to image degradations}
\end{figure}

Figure \ref{fig: Robustness of our method to image degradations} simultaneously illustrates the robustness of the tampering localization and the copyright watermarking to varying image degradations.

Due to the similar principle of embedding, it can be seen that the metrics for the two jobs have similar varying. For two jobs, Gaussian Noise, JPEG Compression, and Brightness Transformation caused a slow decrease; Gaussian Blur and Median Filter caused a relatively great vary when the degradation degree increased; Resize then recover caused a shrinking when the resize ratio was close to 0.1, and we suspect it may be relative with the VAE, which encode images to latent space with the size decreasing 8 times, when the ratio is bigger than $\frac{1}{8}$, the latents have not big difference, giving it a natural resistance to image resize degradations; last, Salt and Pepper Noise can significantly influence two jobs just needing only a little ratio \textemdash more is not necessary.

\section*{Acknowledgements}
This work was supported in part by the Natural Science Foundation of China under Grant 62402469, 62472398, U2336206, by the Fundamental Research Funds for the Central Universities under Grant WK2100000041, and by the Opening Project of MoE Key Laboratory of Information Technology (Sun Yat-sen University) 2024ZD001.



{\small
\bibliographystyle{ieeenat_fullname}
\bibliography{references}
}

\end{document}